\documentclass[article]{revtex4}

\usepackage{graphicx}%
\usepackage{dcolumn}
\usepackage{amsmath}
\usepackage{psfig}
\makeatletter
\def\btt#1{\texttt{\@backslashchar#1}}%
\DeclareRobustCommand\bblash{\btt{\@backslashchar}}%
\makeatother

\begin{document}

\title{Quintessence with O(\textit{N}) Symmetry}

\author{Xin-zhou Li}\email{kychz@shtu.edu.cn}

\author{Jian-gang Hao}
\author{Dao-jun Liu}
\affiliation{ Shanghai United Center for Astrophysics, Shanghai
Normal
University, Shanghai 200234, China\\
East China Institute for Theoretical Physics, Shanghai
200237,China
}%

\date{\today}

\begin{abstract}
ABSTRACT: We study a new class of quintessence models, in which
the scalar field possesses a O(\textit{N}) internal symmetry. We
give a critical condition of instability for the potential against
Q ball formation. We find that the most widely used potentials of
quintessence don't satisfy the above condition, and therefore the
O(\textit{N}) quintessence with these potential will not lead to
the Q ball formation. It is worth noting that O(\textit{N})
quintessence with cosine-type potential is especially interesting
in that the angular contribution is not negligible.
\end{abstract}

\pacs{04.40.-b, 98.80.Cq, 98.80.Es, 98.80.Hw} \maketitle

\vspace{0.4cm} \textbf{1. Introduction} \vspace{0.4cm}

Recent observations on the spectrum of CMB
anisotropies\cite{Bernardis} indicate that about 70 percent of the
total energy in the universe should be hidden as dark
energy\cite{Bahcall}, although these observations should be
extended to even larger red-shifts and smaller angles, as several
missions under preparation will do in the near future.
Observations of type Ia supernova(SNIa)\cite{Perlmutter} shows
that the expansion of universe is accelerating and therefore
requires that the equation-of-state parameter $w=\frac{p}{\rho}$
of the total fluid should be necessarily smaller than
$-\frac{1}{3}$ , where $p$ and $\rho$ are the pressure and energy
density of the fluid in universe respectively. Hitherto, two
possible candidates for dark energy have been suggested . One is
the existence of a cosmological constant and another is a
time-varying scalar field "quintessence"\cite{Ratra,Caldwell},
which has caught much attention ever since its invention. The
quintessence models can be classified into
tracker-type\cite{Steinhardt} and cosine-type\cite{Coble,Kim}. The
tracker-type quintessence has an attractor-like solution which
explains the current dark energy without fine-tuning the initial
condition, while the cosine type one needs tuning of the initial
value of the scalar field although it has been shown that the
fine-tuning problem could be alleviated in the extended
models\cite{Chiba}. This is why most astrophysicist prefer the
tracker quintessence to the cosine one. In fact, even in
tracker-type model, the amplitude of the field and the parameters
in the potential should be carefully adjusted in order to meet the
experimental results. It is worth noting that many theorists tried
to make the potential for tracker-type quintessence arise from
particle physics naturally, even if this issue is really hard
\cite{6}. But the potentials for cosine-type models can arise
rather naturally from particle physics, which attract some
theorists' attention.

 One of the most important observable effects of quintessence is the
 CMB anisotropy and the different types of quintessence
 models will eventually lead to
 different spectrum of the anisotropy. Therefore, more accurate
 measurements\cite{Copeland,Balbi} of the CMB anisotropy (including the forthcoming measurements made by
 MAP\cite{7}, PLANCK\cite{8}) will lay more strict restrictions on
 quintessence models, confirming some of them and abandoning the
 rest.

 A new generalization to quintessence has been
proposed by Boyle et al\cite{Boyle} and Gu and Hwang\cite{Gu}. In
their models, The slowly evolving scalar field of the quintessence
is replaced by a complex scalar field that is spinning in a
circular orbit in a U(1)-symmetric potential. The corresponding
equation of state parameter $w=\frac{p}{\rho}$ for the spinning
quintessence , or spintessence as they call it, has been discussed
and a possible way to distinguish spintessence from other
quintessence scenarios by means of future observations has been
suggested too. It is shown that a complex scalar field can be
regarded as ordinary quintessence if the field does not rotate at
all, and will be spintessence when it rotates rapidly in the
potential.

In this paper, we make a further generalization to spintessence by
replacing the complex scalar field with a \textit{N}-plet scalar
field which is spinning in a O(\textit{N})-symmetric potential.
When one of the angular components is fixed, this O(\textit{N})
quintessence model will reduce to the O(\textit{N}-1) quintessence
model. If the \textit{N}-2 angular components are fixed,
O(\textit{N}) quintessence model will reduce to the U(1)
spintessence model mentioned above. If all angular components are
fixed, O(\textit{N}) quintessence model will reduce to the
quintessence. Especially, the attractor property of the
O(\textit{N})quintessence is still held for some typical
quintessence models and an example is shown in this paper. It is
well known that the scalar field with negative pressure are
generally unstable, the fluctuations may grow rapidly and go
nonlinear to form Q balls\cite{Kusenko, Kasuya, Coleman}. Once the
Q balls are produced, they will act as a dark matter, then the
energy density evolves as $a^{-3}$\cite{Boyle}. In fact, almost
all the charges of the field are absorbed into the produced Q
balls so that there is no homogeneous field left to be dark
energy\cite{Boyle}. This is against the purpose we propose the
quintessence. In this paper, we give a critical condition of
instability for the
 O(\textit{N}) quintessence. Under this condition, the Q ball
formation scenario will occur. Fortunately, we found that the most
widely used potentials for quintessence(Note: some tracker-type
potentials are not stable against Q ball formation at $\lambda
R=O(1)$ epoch) don't satisfy the above condition, and therefore
the O(\textit{N}) quintessence with these potentials are generally
stable.

It is worth noting that this generalization (O(\textit{N})
quintessence) is not very appealing for the tracker-type
quintessence at the late time, for the amplitude of the field will
increase to be very large with the time evolution and the
contribution from the angular component ($\frac{\Omega^2}{a^6
R^3}$, where $R$ is the amplitude of the field) to the equation of
motion becomes negligible as time evolves; but in the early epoch
it will be very important for the contribution from the "angular"
component is not negligible. Furthermore, for the cosine-type
potential, since the amplitude of the quintessence field does not
necessarily increase to be very large to accelerate the expansion
of the universe, and the angular contribution is important for the
whole evolution. Therefore, it is interesting to discuss
O(\textit{N}) quintessence in both tracker-type models and
cosine-type models.

\vspace{0.4cm} \textbf{2. O(N) quintessence}
 \vspace{0.4cm}

We start from the flat Robertson-Walker metric

\begin{equation}
ds^{2}=dt^{2}-a^{2}(t)(dr^{2}+r^{2}d\alpha^{2}+r^{2}sin^{2}\alpha
d\beta^{2})
\end{equation}

\noindent The Lagrangian density for the quintessence with
 O(\textit{N}) symmetry is

\begin{equation}
L_{\Phi}=\frac{1}{2}g^{\mu\nu}(\partial_{\mu}\Phi^{a})(\partial_{\nu}\Phi^{a})-V(|\Phi^{a}|)
\end{equation}

\noindent where $\Phi^{a}$ is the component of the scalar field,
$a=1,2,\cdots,N$. To make it possess a O(\textit{N}) symmetry, we
write it in the following form

\begin{eqnarray}
\Phi^{1}=R(t)\cos\varphi_{1}(t)\hspace{4.2cm}\nonumber\\
\Phi^{2}=R(t)\sin\varphi_{1}(t)\cos\varphi_{2}(t)\hspace{2.85cm}\nonumber\\
\Phi^{3}=R(t)\sin\varphi_{1}(t)\sin\varphi_{2}(t)\cos\varphi_{3}(t)\hspace{1.5cm}\\
\cdots\cdots\hspace{4cm}\nonumber\\
\Phi^{N-1}=R(t)\sin\varphi_{1}(t)\cdots\sin\varphi_{N-2}(t)\cos\varphi_{N-1}(t)\hspace{-0.35cm}\nonumber\\
\Phi^{N}=R(t)\sin\varphi_{1}(t)\cdots\sin\varphi_{N-2}(t)\sin\varphi_{N-1}(t)\nonumber
\end{eqnarray}

\noindent Therefore, we have $|\Phi^{a}|=R$ and assume that the
potential of the O(\textit{N}) quintessence depends only on
\textit{R}. The Lagrangian density then take the following form:

\begin{eqnarray}
L_{\Phi}&=&\frac{1}{2}(\dot{R}^{2}+R^{2}\dot{\varphi}_{1}^{2}+R^{2}\sin^{2}\varphi_{1}\dot{\varphi}_{2}^{2}
+\cdots\hspace{-4cm}\nonumber\\
&+&R^{2}\sin^{2}\varphi_{1}\sin^{2}\varphi_{2}\cdots\sin^{2}\varphi_{N-2}
\dot{\varphi}_{N-1}^{2})-V(R)
\end{eqnarray}

\noindent where the dot denotes the derivative with respect to t.

The action for the universe is :

\begin{equation}
S=\int d^{4}x\sqrt{-g}(-\frac{1}{16\pi G}R_{s}-\rho_{M}+L_{\Phi})
\end{equation}

\noindent where $g$ is the determinant of the metric tensor $
g_{\mu\nu}$ , $G$ is the Gravitational constant, $R_{s}$ is the
Ricci scalar,and $\rho_{M}$ is the non-relativistic matter
density.Using the metric tensor(1) and action (5), we can obtain
the equation of motion for $\varphi_{1}$ , $\cdots $,
$\varphi_{N-1}$ as follows:

\begin{eqnarray}
\ddot{\varphi}_{1}+(2\frac{\dot{R}}{R}+3H)\dot{\varphi}_{1}-\cos\varphi_{1}\sin\varphi_{1}
(\dot{\varphi}^{2}_{2}+\sin^{2}\varphi_{2}\dot{\varphi}^{2}_{3}\hspace{1cm}\nonumber\\+\cdots+\sin^{2}\varphi_{2}
\sin^{2}\varphi_{3}\cdots\sin^{2}\varphi_{N-2}\dot{\varphi}^2_{N-1})=0\hspace{1cm}\\
\ddot{\varphi}_{2}+(2\frac{\dot{R}}{R}+3H+2\cot\varphi_{1}\dot{\varphi}_{1})\dot{\varphi}_{2}
-\cos\varphi_{2}\sin\varphi_{2}(\dot{\varphi}^{2}_{3}+\hspace{0.8cm}\nonumber\\\sin^{2}\varphi_{3}\dot{\varphi}^{2}_{4}
+\cdots+\sin^{2}\varphi_{3}\sin^{2}\varphi_{4}
\cdots\sin^{2}\varphi_{N-2}\dot{\varphi}^{2}_{N-1})=0\hspace{0.5cm}\\
\cdots\cdots\hspace{4cm}\nonumber\\
\ddot{\varphi}_{N-2}+(2\frac{\dot{R}}{R}+3H+2\cot\varphi_{1}\dot{\varphi}_{1}+\cdots+\hspace{2cm}\nonumber\\
2\cot\varphi_{N-3}\dot{\varphi}_{N-3})\dot{\varphi}_{N-2}
-\cos\varphi_{N-2}\sin\varphi_{N-2}\dot{\varphi}^{2}_{N-1}=0\hspace{0.5cm}\\
\ddot{\varphi}_{N-1}+(2\frac{\dot{R}}{R}+3H+2\cot\varphi_{1}\dot{\varphi}_{1}+\cdots+\hspace{2cm}\nonumber\\
2\cot\varphi_{N-2}\dot{\varphi}_{N-2})\dot{\varphi}_{N-1}=0\hspace{1.5cm}
\end{eqnarray}

we have $N-1$ independent first integral of the system of
equations

\begin{eqnarray}
\dot{\varphi}_{1}&=&(\Omega^{2}-\frac{\Omega^{2}_{1}}{\sin^{2}\varphi_{1}})^{\frac{1}{2}}
(a^{3}R^{2})^{-1}\\
\dot{\varphi}_{2}&=&(\Omega^{2}_{1}-\frac{\Omega^{2}_{2}}{\sin^{2}\varphi_{2}})^{\frac{1}{2}}
(a^{3}R^{2}\sin^{2}\varphi_{1})^{-1}
\end{eqnarray}

\begin{eqnarray}
&&\cdots\cdots\nonumber\\
\dot{\varphi}_{N-2}&=&(\Omega^{2}_{N-3}-\frac{\Omega^{2}_{N-2}}{\sin^{2}\varphi_{N-2}})^{\frac{1}{2}}
(a^{3}R^{2}\sin^{2}\varphi_{1}\cdots\nonumber\\\sin^{2}\varphi_{N-3})^{-1}\hspace{-6.3cm}\\
\dot{\varphi}_{N-1}&=&\Omega_{N-2}(a^{3}R^{2}\sin^{2}\varphi_{1}\cdots\sin^{2}\varphi_{N-2})^{-1}
\end{eqnarray}

\noindent where $\Omega$ , $\Omega_{1}$ , $\cdots$ ,
$\Omega_{N-2}$ are $N-1$ independent constants determined by the
initial condition of $\dot{\varphi}_{i}$ , $i=1,2,\cdots,N-1$. The
Einstein equations and the radial equation of scalar fields can be
written as

\begin{equation}
H^{2}=(\frac{\dot{a}}{a})^{2}=\frac{8\pi
G}{3}[\rho_{M}+\rho_{\Phi}]
\end{equation}

\begin{equation}
(\frac{\ddot{a}}{a})=-\frac{8\pi G}{3}[\frac{1}{2}\rho_{M}+2
p_{\Phi}+V(R)]
\end{equation}

\begin{equation}
\ddot{R}+3H\dot{R}-\frac{\Omega^{2}}{a^{6}R^{3}}+\frac{\partial
V(R)}{\partial R}=0
\end{equation}

\noindent where

\begin{equation}
\rho_{\Phi}=\frac{1}{2}(\dot{R}^{2}+\frac{\Omega^{2}}{a^{6}R^{2}})+V(R)
\end{equation}

\begin{equation}
p_{\Phi}=\frac{1}{2}(\dot{R}^{2}+\frac{\Omega^{2}}{a^{6}R^{2}})-V(R)
\end{equation}

\noindent are the energy density and pressure of the $\Phi$ field
respectively , and $H$ is Hubble parameter. The equation-of-state
parameter for the O(\textit{N}) quintessence is

\begin{equation}
w =\frac{p_\Phi}{\rho_\Phi}=
\frac{\dot{R}^{2}+\frac{\Omega^{2}}{a^{6}R^{2}}-2V(R)}{\dot{R}^{2}+
\frac{\Omega^{2}}{a^{6}R^{2}}+2V(R)}
\end{equation}

For the O(\textit{N}) quintessence to accelerate the expansion of
universe , its equation-of-state parameter must satisfy
$w<-\frac{1}{3}$ which is equivalent to

\begin{equation}
\dot{R}^{2}+\frac{\Omega^{2}}{a^{6}R^{2}}<V(R)
\end{equation}

\noindent where the term$\frac{\Omega^{2}}{a^{6}R^{2}}$ comes from
the "total angular motion". The most prominent feature of
O(\textit{N}) quintessence is that it will not reduce to a
cosmological constant even when the $R$ is spatially uniform and
time-independent.

It is worth noting that the introduction of the "angular"
component in O(\textit{N}) quintessence will not change the
attractor property of the dynamical system for the the angular
part will decrease rapidly with the increase of $a$ and $R$. In
order to make it more clear, we will show this property through a
specific model in which, we choose the quintessence potential as
the widely studied tracker potential $V(R)=V_0\exp(-\lambda\kappa
R)$. Following Ref.\cite{copeland2} the system of equations are:

\begin{equation}\label{sys1}
\dot{H}=-\frac{\kappa^2}{2}(\rho_M+p_M+\dot{R}^2 +
\frac{\Omega^2}{a^6R^2})
\end{equation}

\begin{equation}\label{sys2}
\dot{\rho_M}=-3H(\rho_M+p_M)
\end{equation}

\begin{equation}\label{sys3}
\ddot{R}+3H\dot{R}-\frac{\Omega^2}{a^6R^3}-\lambda\kappa V(R)=0
\end{equation}

\begin{equation}\label{sys4}
H^2=\frac{\kappa^2}{3}[\rho_M+\frac{1}{2}(\dot{R}^2+\frac{\Omega^2}{a^6R^2})+V(R)]
\end{equation}

\noindent Here $\rho_M$ and $p_M$ are the energy density and
pressure of the non-relativistic matter and
$p_M=(\gamma-1)\rho_M$, $\gamma$ is a constant, $0\leq
\gamma\leq2$ and $\kappa^2=8\pi G$. Now, introducing the following
variables: $x=\frac{\kappa}{\sqrt{6}H}$,
$y=\frac{\kappa\sqrt{V(R)}}{\sqrt{3}H}$,
$z=\frac{\kappa}{\sqrt{6}H}\frac{\Omega}{a^3R}$,
$\xi=\frac{1}{\kappa R}$ and $N=\log a$, the
Eqs.\ref{sys1}-\ref{sys4} become the following autonomous system:

\begin{equation}\label{auto1}
\frac{dx}{dN}=\frac{3}{2}x[\gamma(1-x^2-y^2-z^2)+2(x^2+z^2)]-(3x-\sqrt{6}z^2\xi-\sqrt{\frac{3}{2}}\lambda
y^2)
\end{equation}

\begin{equation}\label{auto2}
\frac{dy}{dN}=\frac{3}{2}y[\gamma(1-x^2-y^2-z^2)+2(x^2+z^2)]-\sqrt{\frac{3}{2}}\lambda
xy
\end{equation}

\begin{equation}\label{auto1}
\frac{dz}{dN}=-3z-\frac{3}{2}z[\gamma(1-x^2-y^2-z^2)+2(x^2+z^2)]-\sqrt{6}xz\xi
\end{equation}

\begin{equation}\label{auto1}
\frac{d\xi}{dN}=-\sqrt{6}\xi^2x
\end{equation}

It is not difficult to obtain the critical points of the above
autonomous system as: $(x,y,z,\xi)=(0, 0, 0, 0)$, $(-1, 0, 0, 0)$,
$(1, 0, 0, 0)$, $(\frac{\lambda}{\sqrt{6}},
\sqrt{1-\frac{\lambda^2}{6}}, 0, 0)$ and
$(\sqrt{\frac{3}{2}}\frac{\gamma}{\lambda},
\sqrt{\frac{3\gamma(1-\gamma)}{\lambda^2}}, 0, 0)$.  These
critical points will reduce to the case discussed in the ordinary
quintessence model\cite{copeland2} and the "angular" component
will not alter the attractor property because the vanish of
"angular" contribution ($z=0$).

\vspace{0.4cm} \textbf{3. Quintessence without Q ball formation
scenario } \vspace{0.4cm}

 3.1 Condition for Q ball formation \vspace{0.4cm}

As some authors have shown in the context of the Affleck-Dine
baryogenesis \cite{Kusenko,Kasuya}, the formation of Q balls is
very common for the scalar fields with O(\textit{N}) symmetry.
However, this is not favorable for O(\textit{N}) quintessence. In
the following, we will give a condition of instability, under
which the the O(\textit{N}) quintessence model will lead to the
formation of Q balls. For simplicity, we consider only the epoch
that quintessence become dominant.

Follow Ratra and Peebles\cite{Ratra}, we will carry out our
investigation in synchronous gauge and linearize the metric  about
a spatially flat FRW background. The line element is as follow,

\begin{equation}
ds^2=dt^2-a^2(t)(\delta_{ij}-h_{ij})dx^{i}dx^{j}
\end{equation}

\noindent where $h_{ij}$ are the metric fluctuations and $ \mid
h_{ij}\mid<<1 $. For simplicity, we write the first-order
equations of perturbations for the $N=3$ case

\begin{eqnarray}
&&\ddot{(\delta R)}-\frac{1}{a^2}\nabla^2(\delta
R)-[\dot{\varphi_1}^2+\sin^2\varphi_1](\delta R)
+3\frac{\dot{a}}{a}\dot{(\delta R)}\nonumber\\&&+V^{''}(R)(\delta
R)-\frac{1}{2}\dot{h}\dot{R}-2\dot{\varphi_1}R(\dot{\delta
\varphi_1})\nonumber\\
&&-R\sin(2\varphi_1)\dot{\varphi_2}^2(\delta
\varphi_1)-2R\sin^2(\varphi_1)\dot{\varphi_2}(\dot{\delta
\varphi_2})=0
\end{eqnarray}

\begin{eqnarray}
&&\ddot{(\delta \varphi_1)}-\frac{1}{a^2}\nabla^2(\delta
\varphi_1)+3\frac{\dot{a}}{a}\dot{(\delta
\varphi_1)}+2\frac{\dot{R}}{R}(\dot{\delta \varphi_1})\nonumber\\
&&+\dot{\varphi_2}^2\cos(2\varphi_1)(\delta \varphi_1)-\frac{1}{2}\dot{h}\dot{\varphi_1}
-\sin(2\varphi_1)\dot{\varphi_2}(\dot{\delta \varphi_2})\nonumber\\
&&+\frac{2}{R}\dot{\varphi_1}(\dot{\delta
R})-\frac{2}{R^2}\dot{R}\dot{\varphi_1}(\delta R)=0
\end{eqnarray}

\begin{eqnarray}
&&\ddot{(\delta \varphi_2)}-\frac{1}{a^2}\nabla^2(\delta
\varphi_2)+3\frac{\dot{a}}{a}\dot{(\delta
\varphi_2)}+2\frac{\dot{R}}{R}(\dot{\delta \varphi_2})-\frac{1}{2}\dot{h}\dot{\varphi_2}\nonumber\\
&&+2\cot(\varphi_1)[\dot{\varphi_1}(\delta \varphi_2)
+\dot{\varphi_2}(\dot{\delta\varphi_1})]-2\csc^2(\varphi_1)\dot{\varphi_1}\dot{\varphi_2}(\delta
\varphi_1)\nonumber\\
&&+\frac{2}{R}\dot{\varphi_2}(\dot{\delta
R})-\frac{2}{R^2}\dot{R}\dot{\varphi_2}(\delta R)=0
\end{eqnarray}

It is necessary to point out that we here discuss mainly the
stability against Q ball formation, which require that the
fluctuations of the scalar field should not violate the internal
O(\textit{N}) symmetry. That is, only the "radial" part of the
scalar field is perturbed. If the "angular" parts of the scalar
field are perturbed too, then the internal symmetry will not hold
any longer and thus the Q balls will not form even if the
fluctuation is not damped. Therefore, we consider the
perturbation that can preserve the O($N$) symmetry of the scalar
field as following:

\begin{eqnarray}
R(t,\textbf{x})&=&R_0(t)+\delta R(t,\textbf{x}) \nonumber \\
\varphi_{1}(t,\textbf{x})&=&\varphi_{1}(t)\\
\cdots\cdots\nonumber\\
\varphi_{N-1}(t,\textbf{x})&=&\varphi_{N-1}(t)\nonumber
\end{eqnarray}

Then the first order equations of motion for the radial
fluctuation of the scalar field and metric fluctuations are

\begin{eqnarray}\label{dR}
\ddot{(\delta R)}+3H\dot{(\delta R)}-\frac{1}{a^2}\nabla^2(\delta
R)+\frac{\Omega^2}{a^6R_0^4}(\delta R)\nonumber\\
+V^{''}(R_0)(\delta R)-\frac{1}{2}\dot{h}\dot{R_0}=0
\end{eqnarray}

\begin{eqnarray}
\ddot{h}+2H\dot{h}=2\dot{R_0}\dot{(\delta
R)}+2\frac{\Omega^2}{a^6R_0^4}(\delta R)-V^{'}(R_0)(\delta R)
\end{eqnarray}

\begin{eqnarray}
\dot{h}_{,i}-\dot{h}_{ij,j}=\dot{R_{0}}\partial_{i}(\delta R)
\end{eqnarray}

\begin{eqnarray}\label{dhij}
\frac{1}{a^2}(h_{ij,kk}+h_{,ij}-h_{ik,jk}-h_{jk,ik})-3H\dot{h_{ij}}\nonumber\\
-H\dot{h}\delta_{ij}-\ddot{h_{ij}}=\delta_{ij}V^{'}(R_0)(\delta R)
\end{eqnarray}

\noindent where $h$ is the trace of $h_{ij}$. If we choose
$\Omega = 0$, i.e. in the case of $N = 1$ the
Eqs.(\ref{dR})-(\ref{dhij}) will reduce to Ratra-Peebles's case in
the absence of baryonic term\cite{Ratra}. Since the equations of
motion for the metric and scalar fluctuations(Eq.(26) and Eq.(27))
are linear equations, the fluctuations could be taken as the
following form

\begin{equation}
\delta R(t, \textbf{x})=\delta R_0\exp(\alpha
t+i\textbf{k}\textbf{x})
\end{equation}

\begin{equation}
h(t, \textbf{x})=h_0\exp(\alpha t+i\textbf{k}\textbf{x})
\end{equation}

\noindent then for nontrivial $\delta R_0$ and $h_0$, we have

\begin{equation}
\alpha^{2}+(3H-\frac{1}{2}\frac{h_0}{\delta R_0}\dot{R_0})\alpha
-\frac{\Omega^{2}}{a^{6}R_0^{4}}+\frac{k^{2}}{a^{2}}+
\frac{\partial^{2}V}{\partial R_0^{2}}=0
\end{equation}

If $\alpha$ is real and positive, the fluctuations will grow
exponentially , and go nonlinear to form Q balls. Therefore the
instability band for this fluctuation is

\begin{equation}\label{aph}
0<k^{2}<\frac{\Omega^{2}}{a^{4}R^{4}}-\frac{a^{2}\partial^{2}V}{\partial
R^{2} }
\end{equation}

\noindent From Eq.(20) and Eq.(\ref{aph}), it is easily obtained
that the instability band will not exist if the potentials of the
O(\textit{N}) quintessence satisfy

\begin{equation}\label{aph2}
\frac{\partial^{2}V(R)}{\partial R^{2}}>\frac{V(R)}{R^{2}}
\end{equation}

Eq.(\ref{aph2}) is a critical condition for the potential of
O(\textit{N}) quintessence, under which the formation of Q balls
will not occur and thus can be used to determine whether the
potential is "appropriate" for the O(\textit{N}) quintessence
model. In the following , we will show that the two types of
quintessence models all satisfy the above condition.

\vspace{0.4cm}
 \noindent 3.2 The tracker-type quintessence
\vspace{0.4cm}

Tracker-type quintessence models, as previously mentioned , have
an attractor-like solution in a sense that a very wide range of
initial conditions converge to a common evolving track. One can
call the epoch when this attractor-like solution is realized as
"tracking regime", in which the ratio of kinetic energy to
potential energy is almost constant. In fact, the tracker-type
quintessence can also be categorized into two different cases:
case A and case B. For case A, the energy density of the
quintessence is not a significant fraction of the total energy
density of the universe at the early epoch. On the other hand, the
energy density of the quintessence can be significant at early
epoch for the case B.

It is not difficult to prove that the most widely investigated
potentials for case A (such as the inverse power law potential
$V(R)=V_{1}(\frac{R_{0}}{R})^{n}$, which was originally studied
in Ref.\cite{Ratra} and later was investigated in
Ref.\cite{Zlatev}, and the potential
$V(R)=V_{0}[\exp(\frac{R_{0}}{R})-1]$ that has been studied in
Ref.\cite{Skordis}) all satisfy the Eq.(\ref{aph2}). In the
following, we take the potential
$V(R)=V_0\bigg[\exp(\frac{R_0}{R})-1\bigg]$ as an example and
show that it satisfies Eq.(\ref{aph2}). Let

\begin{equation}
f(\frac{R_0}{R})=\frac{R^2}{V_0}\bigg[\frac{\partial
^2V(R)}{\partial R^2}-\frac{V(R)}{R^2}\bigg]
\end{equation}

\noindent and denote $x=\frac{R_0}{R}$, then we have
$f(x)=(x^2+2x-1)e^{x}+1$, when $R\rightarrow \infty$,
$x\rightarrow 0$. Since $f(x)$ is a monotonous increasing function
of $x$, and $f(0)=0$, then for all finite $R$, we have $f(x)>0$,
that is, the potential $V(R)=V_0\bigg[\exp(\frac{R_0}{R})-1\bigg]$
satisfy the condition(\ref{aph2}). Therefore, O(\textit{N})
quintessence models with the above potentials will not lead to Q
ball formation.

As an important example of case B, the potential can be chosen as
\cite{Skordis}

\begin{equation}
V(R)=g(R)e^{-\lambda R}
 \end{equation}

\noindent where $g(R)$ varies more slowly than $e^{-\lambda R}$ in
the tracking regime. The stability condition(\ref{aph2}) is
reduced to

\begin{equation}
\lambda^2-2\lambda\frac{g'}{g}+\frac{g''}{g}>\frac{1}{R^2}
 \end{equation}

\noindent For the simplest case $g(R)=R^2$, we find that
quintessence field is unstable at earlier epoch of universe for
the amplitudes satisfying $2-\sqrt{3}<\lambda R< 2+\sqrt{3}$ in
the tracking regime. This means that case B of tracker
O(\textit{N}) quintessence is unstable during $\lambda R=O(1)$
epoch. But if $\lambda R<2-\sqrt{3}$ or $\lambda R> 2+\sqrt{3}$,
the model is stable against Q ball formation. In fact, the above
result does not depend on the detailed structure of the function
$g(R)$ as long as $g(R)$ is a slowly varying function in the
tracking regime. Thus, the same argument is applicable to case B
with $V(R)=g(R)e^{-\lambda R}$.

 \vspace{0.4cm} \noindent 3.3 The cosine-type quintessence
\vspace{0.4cm}

The potential for Cosine-type quintessence is
\begin{equation}\label{cosV}
V(R)=\Lambda^{4}\bigg[1-\cos(\frac{R}{R_{0}})\bigg]
\end{equation}

\noindent which can be generated if the quintessence is pseudo
Nambu-Goldstone boson field. Firstly, we investigate the evolution
of O(\textit{N}) quintessence in such a potential. There are three
parameters $\Lambda$, $R_0$, and $R_I$ ($R_I$ denotes the initial
value of the field) in the cosine-type potential. Notice that
there is no possible value of $R_I$ consistent with the dark
energy that will account 70 percent energy density for
$\Lambda\lesssim1.9\times10^{-12}\mbox{GeV}$, because the
$\rho_{\Phi}$ is at most $2\Lambda^4$. When $2\Lambda^4$ is close
to $0.7\rho_c$ ($\rho_c$ is the critical density of the present
Universe), $R_I$ should be close to $\pi R_0$ to realize the
relevant value of $0.7\rho_c$. If the slow-roll condition is
satisfied until very recently, the energy density of the
quintessence field becomes dominant and its equation of state
parameter $w$ is close to -1. However, it is not difficult to
obtain that the value of $w$ is very large at the early epoch.
Therefore, there should be a critical value of radius of the
Universe $a_c$. When $a<a_c$, the expansion of the Universe is not
accelerated,

\begin{equation}
\frac{a_c}{a_0}\approx\bigg[\frac{2(1+w_0)}{1-w_0}\bigg]^{\frac{1}{6}}
\end{equation}

\noindent where, $a_0$ and $w_0$ are  the present value of scale
factor of the Universe and $w$ respectively. We can read off this
behavior from the fig1.

\begin{figure}[bp]
\centerline{\psfig{figure=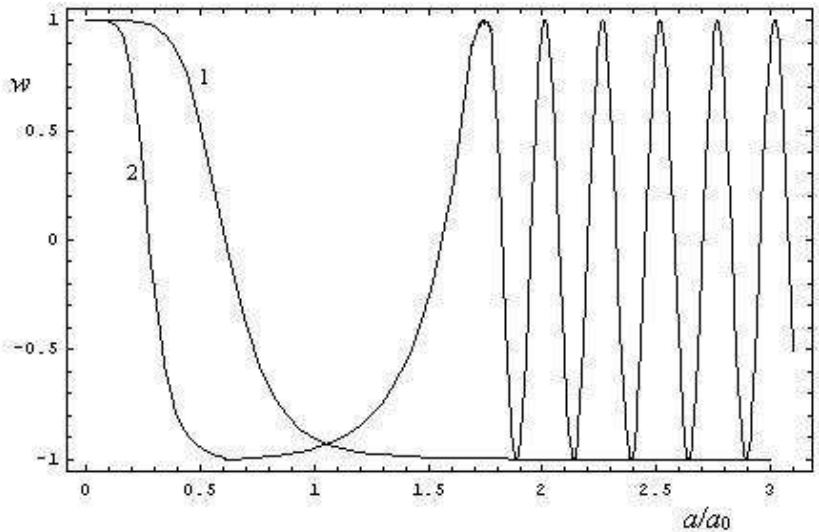,width=5in,height=3.5in}}
\caption{Evolution of the parameter $w$ with respect to
$\frac{a}{a_0}$ for (i) $\Lambda=1.9\times10^{-12}\mbox{GeV}$
(curve 1), and (ii)$4.0\times10^{-12}\mbox{GeV}$ (curve 2)
respectively. Here, we take $R_0=5.0\times10^{17}\mbox{GeV}$,
$w_0=-0.9$ and the Hubble constant
$H_0=65\mbox{km$\cdot$sec$^{-1}\cdot$Mpc$^{-1}$}$.}\label{Fig}
\end{figure}

In contrast, for larger $\Lambda$, $R_I$ smaller than $\pi R_0$ is
possible and the motion of the scalar field becomes important. At
the beginning, the term of total angular momentum is dominant and
the kinetic energy term $\frac{1}{2}\dot{R}^2$ can be neglected.
But, with the expansion of the Universe, the kinetic energy term
and potential energy term become more and more important and
dominant in the end. At the later stage, if we suppose further
that $H<\frac{\Lambda^2}{R_0}$, then the parameter $w$ will
oscillate between -1 and +1. Typical behaviors of the parameter
$w$  as a function of the scale factor are also shown in the fig1.

In this class of models, effective mass of the quintessence field
is always $O( \frac{\Lambda^{2}}{R_{0}})$ and is insensitive to
the amplitude of \textit{R}. If one assumes that (i)the current
quintessence field satisfies the slow-roll condition, i.e. $w
\simeq -1$; (ii)$ \frac{\Lambda^{2}}{R_{0}}$ is not far greater
than the $H$ of the present universe , then the quintessence field
is negligible when $z\gg1$ ($z$ is the redshift). Therefore the
present energy density of the quintessence field is sensitively
depends on the initial amplitude of $R_I$ in the Cosine-type
model. The condition Eq(\ref{aph2})can be rewritten as:

\begin{equation}
\bigg[(\frac{R}{R_{0}})^{2}+1\bigg]\cos(\frac{R}{R_{0}})>1
\end{equation}

\noindent One can find that the stable condition (\ref{aph2}) is
satisfied when  $0<R<\mu_c R_{0}$, where critical value $\mu_c
\approx1.1025$.

Under this circumstance, the variation of the quintessence field
is almost negligible, which make it behave like the cosmological
constant. However, $R_I$ could be smaller than $\pi R_0$ when
$\Lambda$ is very large, and the kinetic energy of the
quintessence becomes important. In particular, the quintessence
will oscillate around the minimum of the potential and thus
satisfy the satability condition(\ref{aph2}) if $\Lambda$ becomes
large enough.That is to say, for the O(\textit{N}) quintessence to
be stable and can accelerate the expansion of the universe, the
amplitude of $R$ is not necessarily goes to be very large and
therefore the angular contribution can not be neglected. In the
$N=1$ case, there are no questions concerning the Q ball
formation. If $\Lambda$ is close to $1.9\times10^{-12} $GeV, then
the $N=1$ is the only choice due to the stability constraint. But
if $\Lambda$ is greater than $1.9\times10^{-12} $GeV, the
O(\textit{N}) model will be possibly stable against Q ball
formation. Since different $\Lambda$ will lead to different CMB
anisotropies which may be detectable in the future satellite
experiments, one can decide whether the cosine-type quintessence
is consist of one scalar field or many fields with an symmetry
constrains by experiments.

 \vspace{0.4cm} \textbf{4. Conclusion and discussion} \vspace{0.4cm}

We have studied a class of new quintessence models, in which the
scalar field possesses a O(\textit{N}) internal symmetry. For the
O(\textit{N}) quintessence models discussed above, the angular
contribution are not negligible at the early epoch of the
universe. While with the time evolving, the potential of the
tracker type O(\textit{N}) quintessence will spill out the angular
momentum because the amplitude $R$ becomes very large. However, we
have shown that the introduction of "angular" component will not
alter the tracker property of the tracker-type quintessence. While
for the cosine-type O(\textit{N}) quintessence, the amplitude of
$R$ does not necessarily goes to be very large and therefore the
angular contribution can not be neglected even in the current
epoch of the universe. We have shown that some tracker-type
O(\textit{N}) quintessence is possibly unstable against Q ball
formation while the others are stable. Therefore, it is very
important to study the cosine-type O(\textit{N}) quintessence and
the corresponding CMB anisotropy caused by it, which we will
discuss in another preparing work.

We also would like to point out that the interaction between
O(\textit{N})quintessence and other scalar fields could be of
certain significance in the formation of cold dark matters. that
is, after the monopoles were produced during the phase transition
in the early universe, the O(2) quintessence might be absorbed
into the monopoles to form a new type of cold stars which are
possible candidates for dark matters. We left the detailed
discussion in another work in preparation.

\vspace{0.8cm}

This work was partially supported by National Nature Science
Foundation of China under Grant No. 19875016, National Doctor
Foundation of China under Grant No. 1999025110, and Foundation of
Shanghai Development for Science and Technology under Grant
No.01JC14035.


\begin{thebibliography}{99}


\bibitem {Bernardis} de Bernardis P et al 2000 Nature {\bf 404} 955\\ Hanany S et al 2000
Astrophys. J. {\bf 545} 1
\bibitem {Bahcall} Bahcall N, Ostriker J P, Perlmutter S and
Steinhardt P J 1999 Science {\bf 284} 1481
\bibitem {Perlmutter} Perlmutter S et al 1999 Astrophys. J. {\bf 517} 565\\ Riess A G et al 1998 Astron. J.
{\bf 116} 1009 astro-ph/9805201
\bibitem {Ratra} Ratra B and Peebles P J 1988 Phys. Rev. {\bf D37} 3406
\bibitem {Caldwell} Caldwell R R, Dave R and Steinhardt P J 1998 Phys. Rev. Lett. {\bf 80} 1582
\bibitem {Steinhardt} Steinhardt P J, Wang L and Zlatev I 1999 Phys. Rev. {\bf D59} 123504 astro-ph/9812313
\bibitem {Coble} Coble K, Dodelson S, Frieman J 1997 Phys. Rev. {\bf D55}
1851
\bibitem {Kim} Kim J E 1999 JHEP {\bf 9905} 022\\Nomura Y, Watari T
 and Yanagida T 2000 Phys. Lett. {\bf B484} 103\\Nomura Y, Watari T
 and Yanagida T 2000 Phys.
Rev. {\bf D61} 105007
\bibitem {Chiba} Chiba T 2001 Phys. Rev. {\bf D64} 103503
\bibitem {6} Masiero A, Pietroni M and Rosati F 2000 Phys. Rev.
{\bf D61} 023504\\ Barreiro T, Copeland E J and Nunes N J 2000
Phys. Rev. {\bf D61} 127301\\ Copeland E J, Nunes N J and Rosati F
2000 Phys. Rev. {\bf D62} 123503
\bibitem {Balbi} Balbi A et al. 2001 Astrophys. J. {\bf 547} L89
\bibitem {Copeland} Corasaniti P S and Copeland E J
astro-ph/0107378 and the references therein
\bibitem {7} MAP homepage, http://map.gsfc.nasa.gov/
\bibitem {8} PLANCK homepage
http://astro.estec.esa.nl/SA-general/Projects/Planck
\bibitem {Boyle} Boyle L A, Caldwell R R and Kamionkowski M
astro-ph/0105318
\bibitem {Gu} Gu Je-An and Hwang W-Y P astro-ph/0105099.
\bibitem {copeland2} Copeland E J, Liddle A R and Wands D 1998 Phys.
Rev. \textbf{D58} 4686
\bibitem {Kusenko} Kusenko A, Shaposhnikov M 1998 Phys. Lett. {\bf B418}
46\\
Kasuya S and Kawasaki M 2000 Phys. Rev. {\bf D61} 041301\\Kasuya S
and Kawasaki M 2000 Phys.Rev. {\bf D62} 023510
\bibitem {Kasuya} Kasuya S astro-ph/0105408
\bibitem {Coleman} Coleman S 1985 Nucl. Phys. {\bf 262} 263
\bibitem {Zlatev}Zlatev I, Wang L  and Steinhardt P J 1999 Phys. Rev. Lett. {\bf 82} 896 astro-ph/9807002
\bibitem {Skordis}Skordis C and Albrecht A astro-ph/0012195
\bibitem {Bento} Bento M C, Bertolami O and  Santos N C astro-ph/0106405

\end{thebibliography}
\end{document}